\documentclass[11pt]{article}
\usepackage{amssymb}
\usepackage{epsfig}
\advance\hoffset by -1.5cm
\advance\voffset by -2cm 
\newtheorem{theorem}{Theorem}[section]
\newtheorem{corollary}[theorem]{Corollary}

\newtheorem{proposition}[theorem]{Proposition}

\newcommand{\beq}{\begin{equation}}
\newcommand{\eeq}{\end{equation}}
\newcommand{\beqa}{\begin{eqnarray}}
\newcommand{\eeqa}{\end{eqnarray}}

\newcommand{\cH}{\mathcal{H}_0}
\newcommand{\cM}{\mathcal{M}}
\newcommand{\cV}{\mathcal{V}}
\newcommand{\cU}{\mathcal{U}}

\begin{document}

\begin{center}
\begin{LARGE}
{\bf Multiple testing via successive subdivision}
\end{LARGE}
\vskip 5mm
\large{\sl Werner Ehm$^a$, J\"urgen Kornmeier$^{a,b}$, Sven Heinrich$^b$}
\end{center}

\begin{abstract}
A sequential multiple testing procedure recently introduced by Heinrich, Bach and Kornmeier allows to ``zoom in" on, and thus identify regions with highly significant departures from null-hypotheses. The purpose of this note is to state a cognate of this procedure in general form and to prove that it controls the familywise error. Two possible applications are briefly indicated. 
\end{abstract}

\vskip 3mm

\section{Introduction}

Often in statistical applications heavy multiple testing is carried out leaving two major questions:\\
Q1: {\em Where} are significant departures from null-hypotheses?\\
Q2: What can be said about the overall error probability of the testing procedure?

In regard to Q2, the classical approach is to control the {\em familywise error,} i.e., to require that the probability of any false rejection is $\le \alpha$, for some $\alpha$ fixed in advance. Such may be achieved using the Bonferroni inequality or, e.g., closed or sequential testing procedures (Marcus et al.~1976, Holm 1979). Particularly when the number of tested hypotheses is large, the desire to avoid any error of the first kind has to be paid by a low test power. Therefore, as an alternative it has been suggested to control instead the {\em false discovery rate} [FDR], i.e., to bound the expected proportion of false rejections among all rejections (Benjamini \& Hochberg 1995). While test power generally is improved with this approach, it does not allow to pin down those tests for which the hypothesis can be safely rejected. Thus when using FDR control one only gets a vague answer to Q1.

There are cases, however, where some tests have very small $p$-values, suggesting a massive violation of the null-hypothesis. Naturally then, one would like to be able to reject precisely those null-hypotheses with guaranteed confidence. A sequential multiple testing procedure designed for such cases has recently been proposed by Heinrich, Bach \& Kornmeier (2008) under the name ``Conquer and Divide" [CaD]. CaD proceeds by successively subdividing the ``search space" and continues testing along each ``search path" until first acceptance of a null-hypothesis, thereby taking advantage of instances where some of the individual tests' $p$-values are very small.

The purpose of this short note is to develop a general, modified version of CaD (also called ``CaD") and to prove that it controls the familywise error. This material appears in Section 2. Section 3 sketches two possible applications. An elaboration of this note is in progress.

\section{The testing procedure}

Consider a rooted tree with vertex set $\cV.$ For definiteness, the tree is supposed to  be ``hanging downward," with the {\em root} $v_0 \in \cV$ on top. Each vertex $v$ splits into its (immediate) {\em descendants,} imagined as lying one layer below $v.$ Let $d(v) \subset \cV$ denote the set of descendants of $v,$ the number 
of which may differ across vertices. The splitting stops at the $L$-th step ($L \ge 1$), such that the vertices of $\cV$ come in $L$ layers below the ($0$-th) root layer. In particular, the tree has {\em depth} $L$ and is {\em complete} in the sense that all branches end at the bottom layer.

With each vertex (a ``location in search space") is associated a testing problem: at every $v \in \cV$ a test of a certain null-hypothesis $\cH(v)$ is carried out whose probability of rejection under $\cH(v)$ is $\le \alpha(v)$. Let us write $\alpha = \alpha(v_0)$ for the test level at the root $v_0$.
The test levels are assumed to satisfy the following \textit{local Bonferroni} condition.

\bigskip\noindent
(LB) \hskip 2mm For every vertex $v \in \cV$ above the $L$-th layer one has $\  \sum\nolimits_{v^\prime \in\, d(v)} \alpha(v^\prime) \le \alpha(v).$

\bigskip\noindent
The proposed multiple testing procedure by successive subdivision may now be described as follows.

\medskip\noindent
[CaD] \textit{Starting at the root $v_0$, keep testing downward each branch of the tree (``search path") as long as the respective null-hypothesis is rejected: stop testing upon first acceptance of a null-hypothesis, and reject all null-hypotheses that have been rejected thus far.}

\medskip\noindent
We will show that the testing procedure is valid, in the sense that its familywise error does not exceed $\alpha$. The familywise error, or probability of an error of the first kind of the {\em procedure} CaD, equals the probability $\pi_1$ that among the hypotheses rejected by CaD there is at least one true (hence falsely rejected) hypothesis.

\begin{proposition}
Under condition {\em (LB)} one has $\pi_1 \le \alpha$.
\end{proposition}

\noindent
{\em Proof.} 
Let $P$ denote the probability measure underlying the observations. Given $P,$ the hypothesis $\cH(v)$ (about $P$) at vertex $v$ is either true or false, independently of the experimental outcome. Thus given $P,$ we get a valued tree by assigning vertex $v$ the truth value $t(v)=0$ if $\cH(v)$ is false, and $t(v)=1$ otherwise. For any vertex $v$ let $U(v)$ denote the set of all vertices $v^\prime \in \cV$ that lie on the (unique) path leading from $v$ up to $v_0$, except for $v$ itself which is excluded.
Let the set $F$ consist of all vertices at which the null-hypothesis is true for the first time, `first' in top-down direction. That is, $F$ comprises all vertices $v \in \cV$ with the following two properties: (i) $t(v^\prime) = 0$ for every $v^\prime \in U(v)$; (ii) $t(v) = 1$. ($F=\{v_0\}$ if $t(v_0)=1.$)

The significance of the set $F$ is the following: (*) if (the application of) CaD happens to produce any error of the first kind (hereafter: ``type I error"), then it also produces a type I error at some vertex $v \in F.$ For suppose that CaD produces a type I error at vertex $v^\ast \in \cV,$ say. If $v^\ast \in F,$ we are done. If $v^\ast \notin F,$ then since $t(v^\ast)=1,$ there exists a first vertex $v$ on the path from $v_0$ down to $v^\ast$ with $t(v)=1,$ that is, there exists $v \in U(v^\ast) \cap F.$ Moreover, the test at $v$ rejects $\cH(v)$ because otherwise the procedure would have stopped at $v,$ leaving no occasion for a type I error to occur at $v^\ast.$ Consequently, a type I error occurs at $v \in F,$ and (*) is proven. But (*) implies
\beqa\label{est1}
\pi_1 & = & P\, [\cH(v)\ \mbox{is rejected for at least one} \ v \in F\, ]\\ & \le & \sum\nolimits_{v \in F} P\, [\cH(v)\ \mbox{is rejected}\,]\nonumber\\ & \le & \sum\nolimits_{v \in F} \alpha(v),\nonumber
\eeqa
whence it suffices to show that
\beq\label{est2}
\sum\nolimits_{v \in F} \alpha(v) \le \alpha.
\eeq
For any complete subtree $\cU$ of $\cV$ let $\rho_\cU$ denote its root vertex. Then (\ref{est2}) is a consequence of the following more general claim:
\beq\label{est3}
\mbox{For every complete subtree $\cU$ of $\cV$, $\ S_\cU := \sum\nolimits_{v \in F \cap \cU} \alpha(v) \le \alpha(\rho_\cU)$.}
\eeq
We argue by induction on the depth $\ell$ of $\cU$ $(0 \le \ell \le L)$. 
The case $\ell = 0$ is trivial (since $\cU$ then consists of its root only), so let $1 \le \ell \, (\le L)$ and suppose that (\ref{est3}) holds for every complete subtree of depth $\ell\!-\!1$. Let $\cU$ be a complete subtree of depth $\ell$. If $F \cap \cU$ is empty or equals $\{\rho_{\cU}\}$, there is nothing to prove. Otherwise let us decompose $\cU\!:$ each descendant $v$ of $\rho_\cU$ represents the root of a complete subtree $\cU(v)$ of $\cU$ of depth $\ell\!-\!1$. Since the vertex sets of all these subtrees are pairwise disjoint, and $\rho_\cU \notin F$ if $F \cap \cU \neq \{\rho_{\cU}\}$, the induction hypothesis and condition (LB) imply
$$
S_{\cU}\, =\, \sum\nolimits_{v\in d(\rho_\cU)} S_{\cU(v)}\, \le\, \sum\nolimits_{v\in d(\rho_\cU)} \alpha(v)\, \le\, \alpha(\rho_\cU).
$$
Thus (\ref{est3}) holds for any complete subtree of depth $\ell$, and the inductive proof is complete.

\bigskip\noindent
{\bf Remarks.} The result immediately generalizes to the case where one has a collection of rooted trees, not necessarily with identical depths, provided the levels of the tests at the roots are controlled by Bonferroni. Note that the significance levels of the tests are moderate initially, and become restrictive only downward the tree. This is in contrast with other sequential procedures, e.g.~Holm's (1979), where the most restrictive tests are carried out first. Note also that no assumption is required about the joint distribution of the test statistics. Finally, control of the familywise error implies that other common error criteria are controlled as well. In fact, domination by the familywise error is guaranteed for any criterion representable as the expected value of a (generally unobservable) random variable with values in $[0,1]$ that assumes the value 0 whenever there is no false rejection. Examples include the false discovery rate and the per comparison error rate (Benjamini \& Hochberg, 1995, p.~291). 

\medskip
A further generalization of the CaD procedure deals with the case where a vertex $v$ may, itself, represent a ``local" multiple testing problem along with an associated testing procedure, $\cM(v)$, rather than just the test of a single hypothesis, $\cH(v)$. The quantity $\alpha(v)$ then has to be interpreted as the familywise error of that testing procedure. 

For example, $\cM(v)$ may stand for the situation where $m = |d(v)|$ null-hypotheses $\cH(v'),\, v' \in d(v)$ are tested using Holm's sequential testing procedure at the level $\alpha(v)$ (familywise). At the next layer, $\cM(v)$ splits into $m$ descendants $\cM(v'),\, v' \in d(v)$, where $\cM(v')$ corresponds to a subdivision of the single hypothesis $\cH(v')$ into a number of further null-hypotheses which, again, are tested using Holm's procedure. Any multiple testing procedure other than Holm's that controls the familywise error can be applied as well. The CaD procedure stops at vertex $v$ if the local procedure associated with $\cM(v)$ accepts {\em at least one} of the single hypotheses $\cH(v')$. Otherwise it continues at {\em all} descendants $\cM(v'),\, v' \in d(v)$. The familywise error $\pi_1$ of the extended CaD procedure is defined as the probability that any of the local testing procedures $\cM(v), \, v \in \cV$ produces a false rejection, which equals the probability that any of the single null-hypotheses $\cH(v')$ is falsely rejected.

\begin{corollary}
Under condition {\em (LB)} the extended {\em CaD} procedure described above satisfies $\pi_1 \le \alpha$.
\end{corollary}

\noindent
{\em Proof. } It suffices to assign truth values as follows: $t(v)=1$ if any of the single hypotheses $\cH(v'),\, v' \in d(v)$ is true, and $t(v)=0$ otherwise. The correspondingly defined set $F$ then retains its original meaning: one readily verifies that if the extended CaD procedure produces a false rejection in the local testing problem $\cM(v^\ast)$, then there is a vertex $v \in F \cap U(v^\ast)$ such that the procedure produces a false rejection in the local testing problem $\cM(v)$. The remainder of the proof is analogous to that of the proposition. 


\medskip
The definition of the extended CaD procedure is chosen such that the original proof carries over easily. Other variants may also be of interest.

\section{Two possible applications}

\noindent
{\bf Analysis of EEG data.}
This is the area CaD was developed for by Heinrich et al.~(2008). In electroencephalographic studies [EEG] one often wants to know where in a time series $\{x(t),\, t \in T\}$ ``something conspicuous" is happening, that is, locate one (or several) time region(s) $C_j \subset T$ showing distinct deviations from the behaviour to be expected under some null-hypothesis $\mathcal{H}_0$. E.g., $\mathcal{H}_0$ may mean ``no systematic departure from zero", $E\, x(t) = 0$ for $t \in T$. With CaD, conspicuous regions are searched for by successively subdividing $T$ into smaller intervals $C_j$ down to a certain level, and testing $\mathcal{H}_0$ restricted to $C_j$ along each subdivision path until first acceptance. Simulations carried out by Heinrich et al.~(2008) suggested that CaD is conservative in the sense of Section 2, and revealed satisfactory power properties.

\medskip\noindent
{\bf Thresholding of wavelet coefficients.}
Nonparametric curve estimation based on thresholding of wavelet coefficients was introduced by Donoho \& Johnstone (1994). As emphasized by Abramovich \& Benjamini (1995), thresholding may be regarded as a multiple testing problem, where an estimated wavelet coefficient $\widehat w_{j,k}$ is kept or set to zero, respectively, in accordance with the outcome of a test of the null-hypothesis that the ``true" coefficient $w_{j,k}=0$. In this context, the above testing procedure could be applied as follows. For $n = 2^{J+1}$ observations, the wavelet coefficients are grouped into resolution levels $j = 1,\ldots,J$ each comprising $2^j$ coefficients $w_{j,k},\, k=1,\ldots,2^j$. They can thus be arranged as a binary tree in which coefficient $w_{j,k}$ ``splits" into $w_{j+1,2k-1}$ and $w_{j+1,2k}$. This splitting corresponds to a halving of time intervals, as is most obvious for the Haar wavelet system. 
The CaD procedure applied with the tests of the hypotheses ``$w_{j,k}=0$" may then be regarded as a method of selecting thresholds for the estimated coefficients $\widehat w_{j,k}$. It differs from related proposals in the literature (e.g., Donoho \& Johnstone (1994), Abramovich \& Benjamini (1995), or, for a different setting, Donoho \& Jin (2008)) in that the threshold is not the same for all coefficients (no matter how adaptive that global value is chosen), but increases with the resolution level. Useful implementations may require modifications of the tests at low resolution levels, in order to avoid too early stopping due to a possible ``averaging out" of wavelet coefficients across longer intervals. The performance of the procedure can be studied along the lines of Abramovich \& Benjamini's (1995) article.

\pagebreak
\section*{References}

\newenvironment{reflist}{\begin{list}{}{\itemsep 0mm \parsep 1mm
\listparindent -7mm \leftmargin 7mm} \item \ }{\end{list}}

\vspace{-7mm}

\begin{reflist}

{\sc Abramovich, F.~\& Benjamini, Y.}~(1995).  Thresholding of wavelet coefficients as multiple hypothesis testing procedure.  In  {\em Wavelets and Statistics}, Ed.~A.~Antoniadis and G.~Oppenheim, Lect.~Notes Statist.~Vol.~103, pp.~5-14. New York: Springer-Verlag.

{\sc Benjamini, Y.~\& Hochberg, Y.}~(1995).  Controlling the false discovery rate: A practical and powerful approach to multiple testing. 
{\em J.~R.~Statist.~Soc.}~B {\bf 57}, 289-300. 

{\sc Donoho, D.~\& Jin, J.}~(2008).  Higher criticism thresholding: Optimal feature selection when useful features are rare and weak.
{\em Proc.~Natl.~Acad.~Sci. (USA)} {\bf 105}, 14790-14795. 

{\sc Donoho, D.~\& Johnstone, I.~M.}~(1994).  Ideal spatial adaptation by wavelet shrinkage.
{\em Biometrika} {\bf 81}, 425-455. 

{\sc Heinrich, S.~P., Bach, M.~\& Kornmeier, J.}~(2008).  Conquer and Divide: A novel approach to spatiotemporal significance testing that accounts for alpha error inflation. {\em Neuroimage} {\bf 41} Suppl. 1, p.~S159.

{\sc Holm, S.}~(1979).  A simple sequentially rejective multiple test procedure. 
{\em Scand.~J.~Statist.} {\bf 6}, 65-70. 

{\sc Marcus, R., Peritz, E.~\& Gabriel, K.R.}~(1976).  On closed testing procedures with special reference to ordered analysis of variance.
{\em Biometrika} {\bf 63}, 655-660.

\end{reflist}

\bigskip\noindent
Author addresses:
\begin{description}
\item[$^a$] Institute for Frontier Areas of Psychology and Mental Health\\ Wilhelmstr.~3a, 79098 Freiburg, Germany
\vspace{-2mm}
\item[$^b$] Department of Ophtalmology, University of Freiburg\\
Killianstr. 5, 79106 Freiburg, Germany
\end{description}

\medskip\noindent
E-mail addresses:

\smallskip\noindent
{\tt ehm@igpp.de\\
kornmeier@igpp.de\\
sven.heinrich@uniklinik-freiburg.de}

\end{document}